\documentclass{emulateapj}

\newcommand{\xte}{{\it RXTE}}

\newcommand{\epcs}{{\rm ergs\,cm^{-2}\,s^{-1}}}

\newcommand{\nmspsmo}{six}
\newcommand{\src}{HETE~J1900.1$-$2455}
\slugcomment{Accepted by ApJL}

\shorttitle{Intermittent pulsations in a millisecond pulsar}
\shortauthors{Galloway et al.}

\begin{document}

\title{% Thermonuclear bursts and
Intermittent pulsations in an
accretion-powered millisecond pulsar} %HETE~J1900.1-2455}

\author{Duncan K. Galloway\altaffilmark{1,2}, %}
  Edward H. Morgan\altaffilmark{3},
  Miriam I. Krauss\altaffilmark{3,4},
  Philip Kaaret\altaffilmark{5}
and
  Deepto Chakrabarty\altaffilmark{3,4}}

\email{D.Galloway@physics.unimelb.edu.au}

\altaffiltext{1}{School of Physics, University of Melbourne, VIC 3010,
  Australia}
\altaffiltext{2}{Centenary Fellow}
\altaffiltext{3}{Kavli Institute for Astrophysics and Space Research,
  Massachusetts Institute of Technology, Cambridge, MA 02139}
\altaffiltext{4}{also Department of Physics, Massachusetts Institute of
  Technology, Cambridge MA 02139 % email: deepto@space.mit.edu}
}
\altaffiltext{4}{Department of Physics and Astronomy, University of Iowa,
  Iowa City, IA 52242}

\begin{abstract}
We describe observations of the seventh accretion-powered millisecond
pulsar, HETE~J1900.1$-$2455 made with the {\it Rossi X-ray Timing
Explorer}\/ during the year of activity that followed its discovery in
2005 June. 
We detected intermittent pulsations at a peak
fractional amplitude of 3\%, but only in the first two
months of the outburst. On three occasions during this time we observed an
abrupt increase in the pulse amplitude, approximately coincident with the
time of a thermonuclear burst, followed by a steady 
decrease on a timescale of $\approx10$~d.
HETE~J1900.1$-$2455 has shown the longest active period by far for any
transient accretion-powered millisecond pulsar, comparable instead to the
outburst cycles for other transient X-ray binaries. Since the last
detection of pulsations, HETE~J1900.1$-$2455 has been indistinguishable
from a low-accretion rate, non-pulsing LMXB; we hypothesize that other,
presently active LMXBs may have also been detectable initially as
millisecond X-ray pulsars.
\end{abstract}


\section{Introduction}

The accretion-powered millisecond pulsars (AMSPs) have
emerged as a distinct class in recent years, beginning with the
discovery of the pulsing nature of SAX~J1808.4$-$3658 in 1998
\cite[]{wij98b,chak98d}. Since then, \nmspsmo\ more AMSPs were discovered during
transient outbursts typically lasting a few weeks
\cite[see][for a review]{wij03a}.
These transient sources verify the evolutionary link between
low-mass X-ray binaries (LMXBs) and recycled millisecond radio pulsars
\cite[]{alpar82,rs82}.  
Extensive observations 
have revealed a number of properties which (until now) appeared
characteristic of the class. The outbursts tend to be of short duration,
typically a few weeks
(but as long as 50~d in XTE~J1814$-$338).
Pulsations are persistently detected at fractional amplitudes of typically
$\sim5$\% rms.  Where thermonuclear bursts are present, oscillations at
the pulsation frequency and roughly the same fractional amplitude are
present throughout \cite[e.g.][]{chak03a,stroh03a}.

\src, the most recently-discovered
AMSP,
remarkably appears to share none of these properties.
This source was discovered when a strong thermonuclear (type-I) burst was
detected 
from a previously unknown source by {\it HETE-II}
\cite[]{vand05a}. A subsequent {\it Rossi X-ray Timing Explorer} (\xte)
observation of the field 
revealed pulsations at 377.3~Hz,
confirming the bursting source as a new pulsar \cite[]{morgan05}. 
The pulse amplitude early in the outburst was as low as
1.5\% rms, fading to below the detection limit following a brief
increase of the persistent flux level 
\cite[]{kaaret05a}. Source activity continued throughout 2005
\cite[]{gal05d} and 2006 (although the source was close to the sun
between 2005 December and 2006 January and hence unobservable).
This is the first 
AMSP that has shown 
``quasi-persistent'' activity.

Here we present analysis of a series of observations over the 2005--06
outburst of \src.
In \S\ref{obs} we describe the observations and analysis procedures.
Our analysis of the pulsation amplitudes and pulse shape
is presented in \S\ref{pulsations}.
We measure the variation of pulse amplitude and arrival time with energy
in \S\ref{pulse_energy}.
Finally, we discuss the implications of our measurements in \S\ref{sec2}.

\section{Observations and Analysis}
\label{obs}

We analysed observations made with 
the Proportional Counter Array \cite[PCA;][]{xte96} aboard \xte, which
consists of five co-aligned proportional counter units (PCUs), sensitive
to photons in the energy range 2--60~keV and with a total effective area
of $\approx6500\ {\rm cm^2}$. Arriving photons are time-tagged to
approximately $1\mu$s, and their energy is measured to a precision of
$<18$\% at 6~keV.

We processed the data using {\sc lheasoft}
\footnote{\url{http://heasarc.gsfc.nasa.gov/docs/software/lheasoft}}
version 5.3.1 (2004 May 19).
For the first observation following the discovery we used a combination of
single-bit and event data modes covering the energy range 3.3--25~keV and
binned on 122~$\mu$s, to measure the properties of the pulsations. For the
remaining observations we used Event mode data (E\_125US\_64M\_0\_1S
configuration) in the energy range 2.5--25~keV binned
every 122~$\mu$s. % to measure the properties of the pulsations.
We used the full-energy range Event mode data from 
observation ID 91015-01-03-01 (2005 June 18)
to measure the variation in pulse properties as a function of energy.
We estimated the background rate using the ``bright-source'' models
appropriate for observations with net count rates $>40\ {\rm
count\,PCU^{-1}\,s^{-1}}$ during PCA gain epoch 5 (from 2000 May
13 onwards)\footnote{\url{http://rxte.gsfc.nasa.gov/docs/xte/pca\_news.html}}.

\begin{deluxetable}{lcll} %{lclccl}
\tablecaption{Type-I X-ray bursts observed from \src
  \label{bursts} }
\tablewidth{0pt}
\tablehead{
  \multicolumn{2}{c}{Start time} & % & \colhead{Peak flux} & \colhead{Fluence}
\\
  \colhead{(UT)} & \colhead{(MJD)} & \colhead{Instrument} &
\colhead{Ref.}
}
\startdata
2005 Jun 14 11:22 & 53535.47361 & {\it HETE-II} % \/ (trigger H3804) % Fregate/WXM/SXC
  & [1,2,3] \\
2005 Jun 17 21:49:10 & 53538.90914 & {\it HETE-II} % & \nodata & \nodata
  & [3] \\ % [3,4] \\
2005 Jun 27 13:54:10 & 53548.57928 & {\it HETE-II} % & \nodata & \nodata
  & [3] \\ % [3,4] \\
2005 Jul 7 13:09:22 & 53558.54891 & {\it HETE-II} & [3] \\ 
2005 Jul 21 23:00:32 & 53572.95871 & {\it HETE-II} & [3] \\
2005 Jul 21 23:00:32 & 53572.95871 & {\it RXTE}/PCA & [4] \\ 
2005 Aug 17 12:19:58 & 53599.51387 & {\it Swift} % \/ (BAT trigger \#150823) % & \nodata & \nodata
  & [5] \\ % [6] \\
2005 Aug 28 15:09:37 & 53610.63167 & {\it Swift} % \/ (BAT trigger \#152451) % & \nodata & \nodata
  & [5] \\ % [6] \\
2005 Sep 24 04:47:10 & 53637.19941 & {\it RXTE}/ASM % & \nodata & \nodata
  & [6] \\ % [7] \\
2006 Mar 20 11:34:06 & 53814.48202 & {\it RXTE}/PCA % (obsid \#92049-01-07-00) % & $9.3\pm0.2$ & $1.524\pm0.009$
  & [4] \\ % [5] \\
\enddata
\tablerefs{1. \cite{vand05a}; 2. \cite{kawai05};
3. \cite{suzuki06};
4. this paper \cite[see also][]{bcatalog}; 
5. C. Markwardt, pers. comm. (2005);
6. R. Remillard, pers. comm. (2005)
}
\end{deluxetable}

\section{Results}
\label{fluxev}

Within the first hundred days of the outburst, the 2.5--25~keV flux varied
by a factor of $\approx2$.
Subsequently the flux stabilised 
at around $8\times10^{-10}\ \epcs$, 
but continued to exhibit
evolution on timescales of $\approx10$~d or so. We observed a number of
more abrupt events;
on MJD~53559 \cite[as also noted by][]{kaaret05b} 
the flux increased by almost 60\% between two observations
separated by just 45~hr, returning back to the earlier level for the
subsequent observation, 49~hr later;
on MJD~53855 the flux dropped by around 40\% between observations
separated by 5.7~d.

Thermonuclear burst activity continued following the 2005 June 14 event
that led to the source discovery \cite[]{vand05a}. Additional bursts were
detected by {\it HETE-II}, {\it Swift}\/ and \xte/ASM (Table \ref{bursts}).
The minimum waiting time between any pair of bursts was 3.44~d; the
maximum (neglecting the gap during which the source was near the sun) was
26.6~d.
Since the persistent flux varied only by a factor of $\approx2$ over this
time, we expect that additional bursts occurred which were not detected
because they fell within intervals when the source was not observable by any
of the satellites.

\begin{figure}
% \epsscale{0.75}	% for submitted version
\epsscale{1.15}	% 2-column version
\plotone{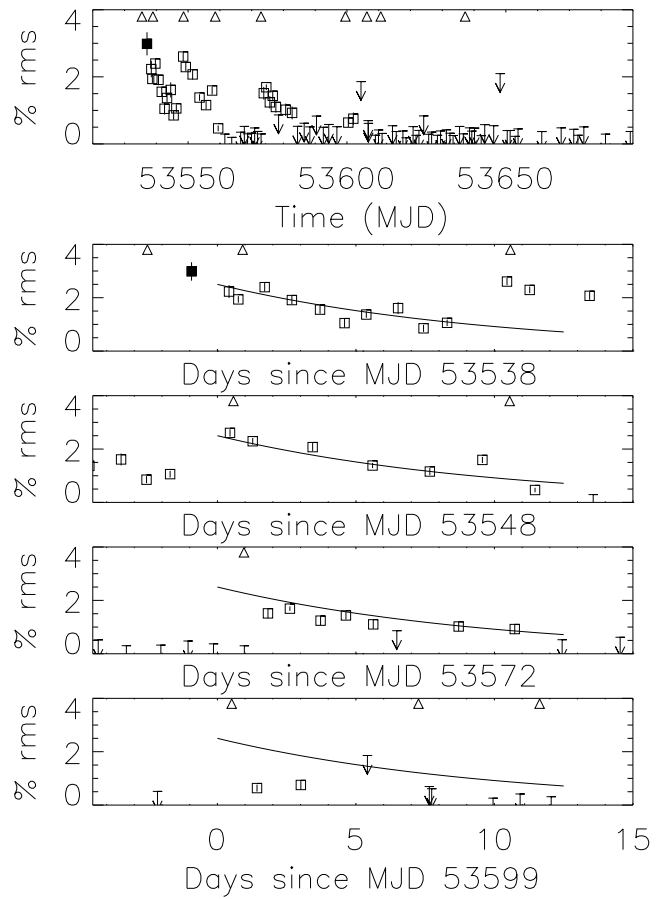}
\caption{Fractional pulse amplitude (\% rms) for \src\/ measured in
the range 2.5--25~keV by \xte\/ during 2005.
Open triangles indicate the times of thermonuclear bursts. 
{\it Top panel}\/ Pulse amplitude history through 
MJD~53690 (2005 November 16). 
The four lower
panels show expanded views of the intervals around the first six bursts
detected. An exponential decay with a maximum of 2.5\% and timescale
$\tau=10$~d is overplotted ({\it solid line}).
For the first \xte\/ observation (on MJD~53537), high-time resolution
modes covering the full PCA band were not available; for that observation
we plot the pulse amplitude in the 3.3--25~keV band ({\it filled symbol}).
\label{rmsvtime} }
\vspace{0.5cm}	% to fix overwriting problem on astro-ph
\end{figure}

 \subsection{Pulse amplitudes and profiles}
 \label{pulsations}

We adjusted the 122~$\mu$s-resolution lightcurves to the solar system
barycenter using the JPL DE200 ephemeris, and corrected for the orbital
motion using the % orbital
parameters of \cite{kaaret05b}. We folded each light-curve on the
pulsar period, and subtracted the expected background rate.
We then fitted the resulting profile to a model consisting of sinusoid
first and second harmonics 
as well as an unpulsed component.
\cite{kaaret05b} noted that the pulsations became undetectable following
the flare on MJD~53559. We detected the pulsations once more beginning
with the observation on MJD~53573 (91059-03-02-00; Fig. \ref{rmsvtime},
top panel), which was not part of the data analysed by \cite{kaaret05b}.
Pulsations were detected through MJD~53582, before becoming undetectable
once more.  We detected the pulsations in just two more observations, the
last on MJD~53602; no detections at $>3\sigma$ have been made since.

We
rarely detected the second harmonic at $>3\sigma$ significance, and
in only one observation where the
first harmonic was present at $>0.7$\% rms. In that observation, the
second harmonic
amplitude was around 24\% of the first (corresponding to a fractional
pulsed amplitude contribution of 0.4\%); in the other detections, the
second harmonic amplitude was between 20 and 60\% of the first.
Since the second harmonic
contributed negligibly to the overall pulse amplitude, we quote the
first harmonic rms amplitude and uncertainties as the pulse amplitudes.

During the initial part of the outburst (through MJD~53560), the pulse
amplitude was typically in the range 0.9--3\% rms (Fig. \ref{rmsvtime}).
The \xte/PCA observations commenced approximately 1.5~d after the discovery
burst on June 14, and initially found the pulse amplitude to be
in the range 2--3\% rms.
Over the following ten days the pulse amplitude decreased on average to
around 1\% rms, until shortly before a third burst was detected by {\it
HETE-II} on MJD~53548 (Table \ref{bursts}). The PCA observation on that
day found the pulse amplitude had returned to close to the maximum
observed previously, at around 2.5\% rms. For the following nine days the
pulse amplitude again decreased to a minimum of $<1$\% rms. Although a
fourth burst was observed on MJD~53558, the 
pulsations became undetectable until 
after the fifth burst, returning at
around 2\% rms. For a third time, the pulse amplitude in the 10~d
following that burst decreased on average, to around 1\% rms and
eventually below the detection limit.
This is further illustrated in the lower panels of Fig.
\ref{rmsvtime}, each of which shows an expanded view of the pulse
amplitude history around the time of the first few bursts. 

We examined the \xte/PCA observations 
closest to the time of bursts in order to estimate precisely when and how
rapidly the pulse amplitude increases.
In at least one case, the pulsations appeared 
strongly again just {\it before}\/ a burst (see Fig. \ref{rmsvtime}, middle
panel). The pulse amplitude for observation 91015-01-04-03 on MJD~53548
was $2.60\pm0.14$\% rms, whereas two days earlier it was $1.06\pm0.12$\%
(having decreased from 
almost 
3\% in the very first PCA observation). % of almost(3.3--25~keV).
Observation 91015-01-04-03 finished at MJD~53548.475, whereas the
nearest burst was detected by {\it HETE-II}\/ at MJD~53548.579, 2.4~hr
{\it after} the end of the observation.

\begin{figure}
% \epsscale{1.0}
\epsscale{1.3}
\plotone{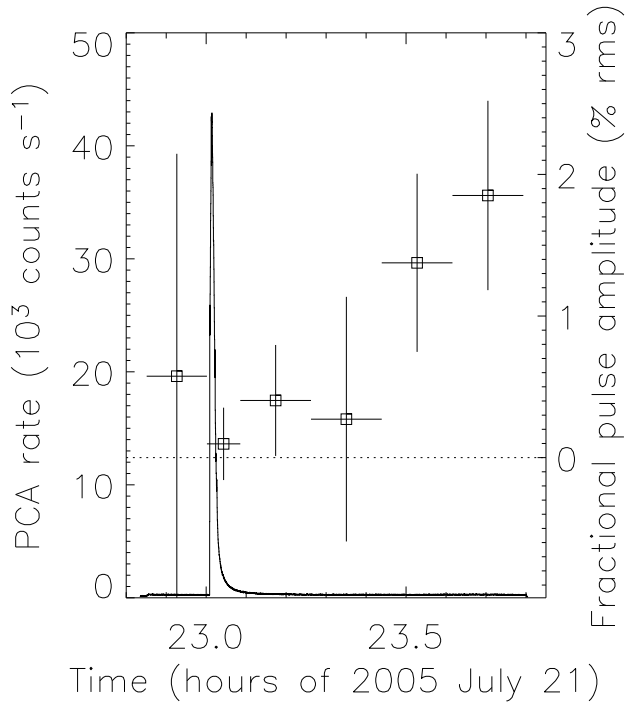}
\caption{Pulse amplitude evolution during the observation of 2005
July 21. The 1-s PCA lightcurve (left-hand $y$-axis) is plotted on the
same $x$-axis as the pulse amplitude measurements ({\it open squares},
right-hand $y$-axis). Error bars indicate the $1\sigma$ uncertainties.
\label{burst} }
\vspace{0.5cm}	% to fix overwriting problem on astro-ph
\end{figure}

In contrast, the first burst detected by the PCA
(on MJD~53572) was not accompanied by the detection of pulsations (the
estimated $3\sigma$ upper limit on the fractional rms pulse amplitude
averaged over the entire observation was
0.57\%) The burst occurred 9.48~min after the start of the observation,
which continued for another 47.5~min 
(Fig. \ref{burst}). We
divided up this observation into six intervals, one each before and during
the burst, and four equally-spaced intervals after, and folded each
lightcurve segment on the pulse period to independently measure the pulse
fraction. We found that the pulse fraction was consistent with zero during
the first four intervals (up to $\approx30$~min after the burst), but in
the last two intervals 
increased to 1.3 and 1.8\%,
respectively.

\subsection{Pulsations as a function of energy}
\label{pulse_energy}

We chose the observation with the highest fractional rms pulse amplitude
to examine the energy dependence of the pulse amplitude and arrival time.
We divided the source spectrum into ten bands to give approximately
the same countrate in each band. We then created light-curves within each
band, binned on 122~$\mu$s, and corrected the times to the solar system
barycenter, as well as correcting for the effects of the binary orbit
using the parameters of \cite{kaaret05b}. We folded each light-curve on
the pulsar period, and measured the amplitude and pulse arrival time by
fitting to a model consisting of a sine curve with a constant offset.

\begin{figure}
% \epsscale{0.8}	% Submitted version only
\epsscale{1.15}	% 2-column version only
\plotone{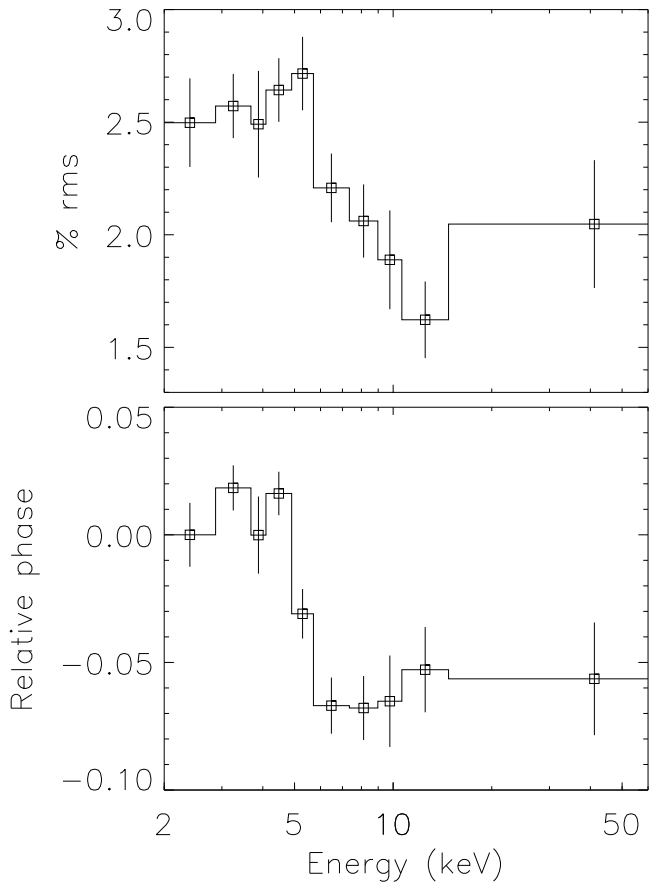}
\caption{Dependence of the pulse profiles and phase with energy for
the observation on 2005 June 18. {\it Top panel}\/ The
fractional rms pulse amplitude as a function of energy. Error bars
indicate the $1\sigma$ uncertainties. {\it Bottom panel}\/ Pulse arrival
time versus energy, relative to the lowest energy channel. A negative value
indicates that the pulse leads the pulse in the first channel.
\label{energy} }
\vspace{0.5cm}	% to fix overwriting problem on astro-ph
\end{figure}

The reduced $\chi^2$ for the fits was consistently $\la1$, indicating no
need for additional harmonics. The pulse fraction was 2.6\% on average
below 6~keV, and then dropped between 6 and 20~keV to as low as 1.6\%
(Fig. \ref{energy}).
There was insufficient signal to 
% determine the behaviour above about
% 20~keV, but the pulse amplitude was consistent with around 2\% in the
% highest band.
further subdivide the energy range above
20~keV and hence test whether the pulse amplitude continued to
evolve as the energy increased. In the mean the pulse amplitude in the
highest band was consistent with around 2\% rms. 
The pulses below 5~keV arrived simultaneously, but the
pulses above this energy arrived earlier, by up to 0.07 cycles. 
Thus, while the pulsations 
had simpler profiles than in the other
AMSPs,
with the first harmonic detected strongly only in one
observation, the energy dependence of the amplitude and pulse arrival time
was similar to that observed in other pulsars \cite[see
e.g.][]{cui98,gp05,fal05}. 

\section{Discussion}
\label{sec2}

\src\ is the most remarkable accretion-powered millisecond pulsar
discovered since the first example, SAX~J1808.4$-$3658. In several
important respects, its behaviour is distinctly different from all other
sources in the class.
First, the source has been active for $>1$~yr, roughly an order of magnitude
longer than any other AMSP. In this respect, the behaviour more closely
resembles transients like MXB~1659$-$298 or KS~1731$-$260, with
outbursts of 2.5 and 12.5~yr duration, respectively \cite[e.g.][]{wij03c}.
For a distance of 5~kpc \cite[]{kawai05}, the mean X-ray flux of \src\
indicates a steady accretion rate of 2--3\%~$\dot{M}_{\rm Edd}$. Although
this is lower than the maximum reached in the outbursts of some other
AMSPs,
the ongoing activity means
a higher time-averaged accretion rate since discovery
\cite[see also ][]{gal06b}.

The ongoing activity may be related to the stellar evolution of the mass
donor. At 83.25~min, the orbital period of \src\ is essentially identical
to the well-known lower limit of $\approx80$~min for a main-sequence mass
donor. The Roche lobe in a binary with a smaller orbital period cannot
accommodate a H-rich donor, and such sources instead
accrete from evolved stars or white dwarfs \cite[]{ps81,nrj86}.
An extended episode of enhanced mass accretion may occur as the
binary evolves through the main-sequence orbital period limit, as the
outer H-rich layers are gradually stripped away leaving only the He-rich
core \cite[see also Fig. 3 of][]{prp02}. 

Second, 
the amplitude of the pulsations 
decreased systematically on a timescale of $~10$~d following 
several of the bursts observed early in the outburst.
\cite{kaaret05b} reported that the pulsations first became undetectable
following a large flare early in the outburst,
likely resulting from a temporary increase in the accretion rate.
Our 
pulse amplitude measurements have revealed that the amplitude had been
decreasing steadily since the burst on MJD~53548, suggesting that the
disappearance of the pulsations after MJD~53559 may have instead been a
consequence of the elapsed time since that burst, and the arrival of the
flare was a coincidence.
Indeed, pulsations only became detectable again
following a subsequent thermonuclear burst which was detected by the PCA.
It remains unclear why the burst just before the flare on MJD~53558, as
well as the bursts observed later in the outburst, did not trigger similar
episodes of high pulse amplitude.

Such pulse amplitude variations have not been reported in any other
AMSP; we note that bursts have been detected in two
others, SAX~J1808.4$-$3658 \cite[]{chak03a} and XTE~J1814$-$314
\cite[]{stroh03a}.
It is particularly intriguing that the onset of the
pulsations can apparently occur up to 2.4~hr {\it before} or 0.5~hr {\it
after} the actual time of burst ignition (or not at all). That the
pulsations may appear prior to the burst seems to rule out oscillations
which are directly triggered by the bursts, 
although some dependence of the pulsation mechanism on the properties of
the envelope is indicated.

The third aspect in which \src\ behaves distinctly differently from the
other six known AMSPs
is that
the pulsations in \src\ were present only in the first few months
of the outburst.
While the amplitude of the pulsations in the other AMSPs may
vary throughout the outburst, they are always detectable
except at the end of the outburst when the
source flux has dropped essentially to the background level.
It is an open question as to what physics prevents
persistent millisecond pulsations from being detected in most 
LMXBs containing neutron stars. 
That \src\ apparently makes a transition from persistently-pulsing
(albeit at rather low amplitude) to a non-pulsing, quasi-persistent source
within a timescale of a month indicates that the visibility of the
pulsations is determined in the outer layers of the neutron star
envelope, since this timescale is far too short to affect the % thermal
properties of the core.
One possibility that has been discussed theoretically is that magnetic
channeling of the flow (which is thought to give rise to the pulsations)
will cease once the magnetic field is ``buried'' by the accreted material
\cite[e.g.][]{czb01,payne05}. The cessation of pulsations 70~d after the
beginning of the outburst in \src\ suggests that only $10^{-10}\ M_\odot$,
or equivalently a sustained rate of $\approx2$\%~$\dot{M}_{\rm Edd}$ is
necessary to bury the field. 
We hypothesize that other, presently-active low-$\dot{M}$ LMXBs may have
also been detectable as millisecond X-ray pulsars when they first became
active.

\acknowledgments

We thank Craig Markwardt, Nat Butler, and Ron Remillard for supplying the
times of bursts observed by {\it Swift}, {\it HETE-II}, and {\it
RXTE}/ASM, respectively.
This work was supported in part by
the NASA Long Term Space Astrophysics program under grant NAG 5-9184.

% \bibliography{all}
% \bibliographystyle{apj}

\end{document}